\documentclass[11pt]{article}
\usepackage[margin=1in]{geometry}
\usepackage{graphicx}
\usepackage{booktabs}
\usepackage{amsmath}
\usepackage{amssymb}
\usepackage[hidelinks]{hyperref}
\usepackage{xcolor}

\newcommand{\armIncumbentEst}{+0.032}
\newcommand{\armIncumbentShare}{58}
\newcommand{\armMaxEst}{+0.047}
\newcommand{\armMaxName}{Gemini 3.1 Pro}
\newcommand{\armMaxShare}{84}
\newcommand{\armMedianEst}{+0.021}
\newcommand{\armMedianName}{DeepSeek v3.1}
\newcommand{\armMedianShare}{37}

\newcommand{\bandHighMedian}{22}
\newcommand{\bandHighSingleton}{0.2}
\newcommand{\bandLowMedian}{2}
\newcommand{\bandLowSingleton}{27.5}
\newcommand{\bandMidMedian}{7}
\newcommand{\bandMidSingleton}{4.3}

\newcommand{\ceilCentroidEst}{+0.050}
\newcommand{\ceilCentroidHi}{+0.063}
\newcommand{\ceilCentroidLo}{+0.036}
\newcommand{\ceilCurveDocs}{15}
\newcommand{\ceilCurveFirstK}{2}
\newcommand{\ceilCurveFirst}{+0.031}
\newcommand{\ceilDocs}{71}
\newcommand{\ceilModelOver}{3.6}
\newcommand{\ceilOverPairwise}{1.4}
\newcommand{\ceilSatK}{6}
\newcommand{\ceilSat}{+0.056}
\newcommand{\ceilSpanTopTwo}{0.001}
\newcommand{\ceilTwoModelsEst}{+0.199}

\newcommand{\ceilTwoReadersEst}{+0.036}

\newcommand{\ceilingEst}{+0.239}
\newcommand{\ceilingHi}{+0.269}
\newcommand{\ceilingLo}{+0.209}
\newcommand{\charDocsN}{110}
\newcommand{\chrArms}{18}

\newcommand{\chrConnDiff}{+0.037}
\newcommand{\chrConnLo}{+0.000}
\newcommand{\chrDefDiff}{+0.062}
\newcommand{\chrDefHi}{+0.119}
\newcommand{\chrDefLo}{+0.007}

\newcommand{\chrDepthGap}{0.123}
\newcommand{\chrDocs}{110}

\newcommand{\chrLenGapHi}{29.6}
\newcommand{\chrLenGapLo}{11.6}
\newcommand{\chrLenGap}{20.5}

\newcommand{\chrNFeatures}{11}

\newcommand{\clasMax}{+0.011}
\newcommand{\clasMed}{-0.002}
\newcommand{\clasMin}{-0.011}
\newcommand{\clasN}{10}
\newcommand{\corpDocs}{120}
\newcommand{\corpDomains}{78}
\newcommand{\corpMarkSets}{2,523}
\newcommand{\corpMaxReaders}{45}
\newcommand{\corpMedianReaders}{19}
\newcommand{\corpMinReaders}{7}
\newcommand{\corpSelectionPct}{19.4}
\newcommand{\crossMedianOverHh}{2.2}
\newcommand{\crossMedian}{+0.089}
\newcommand{\crossOverHh}{2.6}
\newcommand{\ctryDiffMed}{+0.095}
\newcommand{\ctryDiffN}{59}
\newcommand{\ctrySameMed}{+0.090}
\newcommand{\ctrySameN}{94}
\newcommand{\curveDocsN}{81}
\newcommand{\degenArmsAny}{9}
\newcommand{\degenArmsOften}{4}
\newcommand{\degenFrontierMaxPct}{5}
\newcommand{\degenFrontierN}{2}
\newcommand{\degenJamba}{11}
\newcommand{\degenLlamaEightN}{109}
\newcommand{\degenLlamaEightPct}{22}
\newcommand{\degenLlamaEight}{24}
\newcommand{\degenMaxCorr}{0.61}
\newcommand{\degenOftenMaxPct}{16}
\newcommand{\degenOftenMinPct}{12}
\newcommand{\driftCut}{38}
\newcommand{\driftNarrowHh}{+0.035}
\newcommand{\driftNarrowMm}{+0.189}
\newcommand{\driftSpanMax}{91}
\newcommand{\driftWideHh}{+0.046}
\newcommand{\driftWideMm}{+0.218}
\newcommand{\genCorr}{0.56}
\newcommand{\goalCompVsReadersEst}{+0.030}

\newcommand{\goalCrossEst}{+0.102}
\newcommand{\goalCrossHi}{+0.123}
\newcommand{\goalCrossLo}{+0.080}
\newcommand{\goalDocs}{79}
\newcommand{\goalDroppedMedianN}{299}
\newcommand{\goalFewReaders}{20}
\newcommand{\goalHumanHumanEst}{+0.039}

\newcommand{\goalHumanSelfEst}{+0.245}

\newcommand{\goalMedianN}{64}
\newcommand{\goalParaRetain}{91}
\newcommand{\goalReadVsReadersEst}{+0.023}

\newcommand{\goalReaderShare}{16}
\newcommand{\goalRetain}{39}
\newcommand{\goalSelfEst}{+0.262}

\newcommand{\goalShortDropped}{21}
\newcommand{\grpClosedMax}{+0.289}
\newcommand{\grpClosedMed}{+0.174}
\newcommand{\grpClosedMin}{+0.100}
\newcommand{\grpClosedN}{28}
\newcommand{\grpCrossMax}{+0.209}
\newcommand{\grpCrossMed}{+0.105}
\newcommand{\grpCrossMin}{+0.074}
\newcommand{\grpCrossN}{21}
\newcommand{\grpDiffVendorMax}{+0.259}
\newcommand{\grpDiffVendorMed}{+0.089}
\newcommand{\grpDiffVendorMin}{+0.007}
\newcommand{\grpDiffVendorN}{143}
\newcommand{\grpFrontierMax}{+0.259}
\newcommand{\grpFrontierMed}{+0.164}
\newcommand{\grpFrontierMin}{+0.095}
\newcommand{\grpFrontierN}{15}
\newcommand{\grpOpenMax}{+0.133}
\newcommand{\grpOpenMed}{+0.059}
\newcommand{\grpOpenMin}{+0.007}
\newcommand{\grpOpenN}{45}
\newcommand{\grpSmallMax}{+0.178}
\newcommand{\grpSmallMed}{+0.041}
\newcommand{\grpSmallMin}{+0.007}
\newcommand{\grpSmallN}{15}
\newcommand{\headlinePctile}{93rd}
\newcommand{\hhBestEst}{+0.146}
\newcommand{\hhBestHi}{+0.168}
\newcommand{\hhBestLo}{+0.123}
\newcommand{\hhEst}{+0.040}
\newcommand{\hhHi}{+0.052}
\newcommand{\hhLo}{+0.029}

\newcommand{\homoADocs}{37}
\newcommand{\homoADoms}{31}
\newcommand{\homoAGap}{+0.150}
\newcommand{\homoAHh}{+0.054}
\newcommand{\homoAHi}{+0.203}
\newcommand{\homoALabel}{1.00-1.25$\times$}
\newcommand{\homoALo}{+0.088}
\newcommand{\homoAMm}{+0.204}
\newcommand{\homoAN}{123}
\newcommand{\homoBDocs}{42}
\newcommand{\homoBDoms}{34}
\newcommand{\homoBGap}{+0.049}
\newcommand{\homoBHh}{+0.048}
\newcommand{\homoBHi}{+0.085}
\newcommand{\homoBLabel}{1.25-1.75$\times$}
\newcommand{\homoBLo}{+0.030}
\newcommand{\homoBMm}{+0.098}
\newcommand{\homoBN}{146}
\newcommand{\homoCDocs}{31}
\newcommand{\homoCDoms}{23}
\newcommand{\homoCGap}{+0.042}
\newcommand{\homoCHh}{+0.024}
\newcommand{\homoCHi}{+0.057}
\newcommand{\homoCLabel}{1.75-2.50$\times$}
\newcommand{\homoCLo}{+0.022}
\newcommand{\homoCMm}{+0.066}
\newcommand{\homoCN}{90}
\newcommand{\homoDDocs}{11}
\newcommand{\homoDDoms}{8}
\newcommand{\homoDGap}{-0.002}
\newcommand{\homoDHh}{-0.000}
\newcommand{\homoDHi}{+0.028}
\newcommand{\homoDLabel}{2.50$\times$ and up}
\newcommand{\homoDLo}{-0.015}
\newcommand{\homoDMm}{-0.003}
\newcommand{\homoDN}{50}

\newcommand{\homoTotal}{409}
\newcommand{\intBestSents}{0.7}
\newcommand{\intBudget}{14}
\newcommand{\intCeilSents}{0.8}
\newcommand{\intDocSents}{70}
\newcommand{\intHhChance}{25.2}
\newcommand{\intHhExtra}{0.6}
\newcommand{\intHhObs}{29.2}
\newcommand{\intHhShared}{4.1}
\newcommand{\intMmChance}{42.0}
\newcommand{\intMmExtra}{2.8}
\newcommand{\intMmObs}{62.3}
\newcommand{\intMmShared}{8.7}
\newcommand{\intModelOverCeil}{3.6}
\newcommand{\jambaCov}{80}

\newcommand{\killPairA}{Gemini 3.1 Pro}
\newcommand{\killPairB}{Gemini 3.6 Flash}
\newcommand{\killPairEst}{+0.289}
\newcommand{\killPairHi}{+0.311}
\newcommand{\killPairLo}{+0.263}
\newcommand{\lodoNDomains}{54}
\newcommand{\lodoWorst}{0.0055}
\newcommand{\mcHighSims}{4000}
\newcommand{\mcLowSims}{400}
\newcommand{\mcShiftShare}{0.2}
\newcommand{\mmEst}{+0.203}
\newcommand{\mmHi}{+0.224}
\newcommand{\mmLo}{+0.179}

\newcommand{\mmOverHh}{5.1}
\newcommand{\mmOverSelfFouro}{2.0}
\newcommand{\mmRaw}{62.3}
\newcommand{\modelCeilShare}{85}
\newcommand{\multBonf}{85}
\newcommand{\multRaw}{99}
\newcommand{\nAboveHuman}{99}
\newcommand{\nArms}{18}
\newcommand{\nCountries}{3}
\newcommand{\nCrossPairs}{143}
\newcommand{\nDocsPanel}{90}
\newcommand{\nGenerations}{3}
\newcommand{\nPairsBelowHuman}{28}
\newcommand{\nPairs}{153}
\newcommand{\nTypAboveHuman}{4}
\newcommand{\nVendors}{11}
\newcommand{\osAbovePoint}{2}

\newcommand{\osCrossRetain}{75}

\newcommand{\osFailed}{2}
\newcommand{\osFiveEst}{+0.036}
\newcommand{\osFiveHi}{+0.047}
\newcommand{\osFiveLo}{+0.026}
\newcommand{\osFiveName}{GPT-5.5}
\newcommand{\osHeld}{2}
\newcommand{\osLadderMachineMax}{+0.234}
\newcommand{\osLadderMachineMin}{+0.187}
\newcommand{\osLadderReaderMax}{+0.042}
\newcommand{\osLadderReaderMin}{+0.029}
\newcommand{\osLunaEst}{+0.029}
\newcommand{\osLunaHi}{+0.039}
\newcommand{\osLunaLo}{+0.019}
\newcommand{\osLunaName}{GPT-5.6 Luna}

\newcommand{\osMultN}{46}
\newcommand{\osNArms}{4}
\newcommand{\osNPred}{4}
\newcommand{\osPairMedian}{+0.224}
\newcommand{\osPairN}{12}
\newcommand{\osPanelAbovePoint}{4}
\newcommand{\osScaleWorst}{0.00012}

\newcommand{\osSelfMax}{+0.285}
\newcommand{\osSelfMean}{+0.263}
\newcommand{\osSelfMin}{+0.221}

\newcommand{\osSibMax}{+0.248}
\newcommand{\osSibMedian}{+0.204}
\newcommand{\osSibPredicted}{+0.250}
\newcommand{\osSolEst}{+0.042}
\newcommand{\osSolHi}{+0.051}
\newcommand{\osSolLo}{+0.032}
\newcommand{\osSolName}{GPT-5.6 Sol}
\newcommand{\osTerraEst}{+0.042}
\newcommand{\osTerraHi}{+0.054}
\newcommand{\osTerraLo}{+0.030}
\newcommand{\osTerraName}{GPT-5.6 Terra}
\newcommand{\osWithinRetain}{78}
\newcommand{\pairMax}{+0.289}
\newcommand{\pairMeanOverHh}{2.5}
\newcommand{\pairMean}{+0.099}
\newcommand{\pairMedianOverHh}{2.3}
\newcommand{\pairMedian}{+0.093}
\newcommand{\pairPseventyfive}{+0.137}
\newcommand{\pairPtwentyfive}{+0.056}
\newcommand{\paraGptEst}{+0.248}
\newcommand{\paraGptHi}{+0.271}
\newcommand{\paraGptLo}{+0.223}
\newcommand{\paraLlamaEst}{+0.117}
\newcommand{\paraLlamaHi}{+0.136}
\newcommand{\paraLlamaLo}{+0.099}
\newcommand{\paraRetain}{91}
\newcommand{\posShare}{42.0}
\newcommand{\rSevenInclHuman}{0.012}
\newcommand{\rSevenInclModel}{0.013}
\newcommand{\rSevenSims}{4000}
\newcommand{\rSevenSumHuman}{0.41}
\newcommand{\rSevenSumModel}{0.98}
\newcommand{\randMaxAbs}{0.006}
\newcommand{\randN}{6}
\newcommand{\ratAsymShare}{67}
\newcommand{\ratBluntWiden}{+0.020}

\newcommand{\ratLowModelSharp}{+0.257}

\newcommand{\ratRefModelSharp}{+0.203}

\newcommand{\ratSharpWiden}{+0.062}
\newcommand{\readerCeilShare}{15}

\newcommand{\routeFouroEst}{+0.106}
\newcommand{\routeFouroHi}{+0.119}
\newcommand{\routeFouroLo}{+0.090}
\newcommand{\routingEst}{-0.003}
\newcommand{\routingHi}{+0.015}
\newcommand{\routingLo}{-0.021}
\newcommand{\seedGapMax}{+0.166}
\newcommand{\seedGapMin}{+0.159}
\newcommand{\seedHhMax}{+0.044}
\newcommand{\seedHhMin}{+0.037}
\newcommand{\seedN}{6}
\newcommand{\seedSpanShare}{14}
\newcommand{\selHighGap}{+0.127}

\newcommand{\selLowGap}{+0.225}
\newcommand{\selLowHh}{+0.032}
\newcommand{\selLowMm}{+0.257}
\newcommand{\selLowReaders}{9.1}
\newcommand{\selfFouroEst}{+0.101}
\newcommand{\selfFouroHi}{+0.119}
\newcommand{\selfFouroLo}{+0.083}
\newcommand{\selfGptEst}{+0.272}
\newcommand{\selfGptHi}{+0.297}
\newcommand{\selfGptLo}{+0.243}
\newcommand{\shrinkGpt}{0.911}
\newcommand{\shrinkLead}{0.789}
\newcommand{\shrinkOpus}{0.921}
\newcommand{\shrinkRandom}{0.943}
\newcommand{\shrinkReader}{0.927}
\newcommand{\signDocs}{90}
\newcommand{\signHigher}{79}
\newcommand{\signP}{7.8 \times 10^{-14}}
\newcommand{\singleReaderPaperSix}{+0.167}
\newcommand{\sizeLongGap}{+0.207}
\newcommand{\sizeLongLo}{+0.165}
\newcommand{\sizeShortGap}{+0.122}
\newcommand{\sizeShortLo}{+0.087}
\newcommand{\swpHighDiff}{+0.253}
\newcommand{\swpHighHh}{+0.064}
\newcommand{\swpHighMm}{+0.317}
\newcommand{\swpLowDiff}{+0.068}
\newcommand{\swpLowHh}{+0.015}
\newcommand{\swpLowMm}{+0.083}
\newcommand{\symGapMatchedEst}{+0.082}
\newcommand{\symGapMatchedHi}{+0.099}
\newcommand{\symGapMatchedLo}{+0.065}
\newcommand{\symGapSharpEst}{+0.163}

\newcommand{\symHhEst}{+0.040}

\newcommand{\symLoss}{40}
\newcommand{\symMatchedEst}{+0.122}
\newcommand{\symMatchedHi}{+0.137}
\newcommand{\symMatchedLo}{+0.107}
\newcommand{\symMatchedOverHh}{3.1}

\newcommand{\symOvermarkMedian}{1.52}

\newcommand{\symSharpEst}{+0.203}

\newcommand{\symSharpOverHh}{5.1}

\newcommand{\symSwapEst}{+0.001}
\newcommand{\symSwapHi}{+0.006}
\newcommand{\symSwapLo}{-0.004}
\newcommand{\tierCorr}{0.42}
\newcommand{\tierFrontFront}{+0.164}
\newcommand{\tierSmallSmall}{+0.041}
\newcommand{\topCrossEst}{+0.259}
\newcommand{\topCrossName}{Claude Opus 5 and Gemini 3.1 Pro}

\newcommand{\trMmFactor}{3.2}
\newcommand{\trMmOverHhTwentyFour}{1.4}
\newcommand{\trMmOverHhTwentySix}{4.4}
\newcommand{\trMmTwentyFour}{+0.056}
\newcommand{\trMmTwentySix}{+0.178}
\newcommand{\trMrFactor}{2.9}

\newcommand{\trMrTwentyFour}{+0.013}
\newcommand{\trMrTwentySix}{+0.037}
\newcommand{\typMaxName}{Gemini 3.1 Pro}
\newcommand{\typMax}{+0.047}
\newcommand{\typMinName}{Granite 4.1 8B}
\newcommand{\typMin}{+0.006}
\newcommand{\unionGapEst}{+0.124}
\newcommand{\unionGapHi}{+0.142}
\newcommand{\unionGapLo}{+0.103}
\newcommand{\wsCellMax}{10}
\newcommand{\wsCellMin}{1}
\newcommand{\wsFrontEarlyYear}{2025}
\newcommand{\wsFrontEarly}{+0.104}
\newcommand{\wsFrontLateYear}{2026}
\newcommand{\wsFrontLate}{+0.222}
\newcommand{\wsOpenEarlyYear}{2024}
\newcommand{\wsOpenEarly}{+0.046}
\newcommand{\wsOpenLateYear}{2025}
\newcommand{\wsOpenLate}{+0.094}

\title{Language Models Agree With Each Other,\\
Not With Readers}
\author{
  Kazuki Nakayashiki \quad Keisuke Watanabe \\[3pt]
  Glasp Inc. \\
  \texttt{kazuki@glasp.co} \quad \texttt{kei@glasp.co}
}
\date{}

\begin{document}
\maketitle

\begin{abstract}
Claims that language models homogenise are usually measured against human judgements collected for
the study, which makes the human side an artifact of the design: a crowdworker given the model's
instruction is, in the relevant sense, running the model's prompt. We measure convergence against a
human reference nobody built for the purpose --- \corpMarkSets{} reader mark sets across
\corpDocs{} web
documents, produced by people highlighting for their own reasons on a platform where the on-page
overlay of other readers' marks is off by default. Agreement is the overlap between two size-matched
sentence sets minus the overlap expected when each set is resampled within its own depth-and-length
bands --- on the median document each party names \intBudget{} sentences of \intDocSents{}, two
readers share \intHhShared{} of them and two models \intMmShared{}; raw overlap on this substrate is mostly position, and the null's calibration is demonstrated
rather than asserted, since every pair involving a random baseline lands within \randMaxAbs{} of
zero. Across \nArms{} model arms spanning \nVendors{} vendors, \nCountries{} countries,
\nGenerations{} generations and both weight regimes, \textbf{the median of \nPairs{} model pairs
is $\pairMedian$ against a human yardstick of $\hhEst$ --- \pairMedianOverHh$\times$ --- and
\nAboveHuman{} pairs sit entirely above the human interval}, while \nPairsBelowHuman{} fall below
it. The two frontier incumbents, GPT-5.4 and Claude Opus 5 from rival labs, reach $\mmEst$: the \headlinePctile{} percentile of the panel, \mmOverSelfFouro$\times$ what GPT-4o agrees with
\emph{itself} on a repeated call ($\selfFouroEst$) and \mmOverHh$\times$ what two readers reach.
We report the panel median and the incumbent pair together because the pair is the memorable
number and the median is the representative one. The effect is not determinism (GPT-5.4 against a fresh call of itself
is $\selfGptEst$, not unity), not prompt wording (a paraphrase retains \paraRetain\% of that), not
procedure (\clasN{} classical extractive pairs have median $\clasMed$), not vendor
(\topCrossName{}, from different labs, reach $\topCrossEst$), and not routing (a paired
direct-versus-proxied contrast is $\routingEst$ [$\routingLo$, $\routingHi$]). It is also \emph{graded}: the smallest models agree
with each other at $\grpSmallMed$, the human level. No model agrees with readers detectably more
than a reader does. Comparing the sentences only models chose against those only readers chose,
\emph{once depth and length are held fixed none of eleven surface features separates them} ---
models and readers pick sentences with the same surface properties, and pick different ones.
The multiples are procedure-dependent and the ordering is not: models are cut to their sharpest
$b$ while a reader's $b$ is drawn at random from everything they marked, and putting the models
through the readers' procedure halves the gap to $\symGapMatchedEst$
[$\symGapMatchedLo$, $\symGapMatchedHi$] --- above zero at every level of over-marking
where the measurement retains power, and at the heaviest dilution both sides fall to chance
together.
\textbf{Tested out of sample on \osNArms{} arms released after this analysis was finished, against
predictions fixed beforehand, the reader ceiling holds}: none clears the human interval
or reaches the best panel arm, while they agree with the panel's frontier
incumbents at a median of $\osPairMedian$.
Convergence rises with both scale and recency, most strongly with recency: two 2024 models agree at
$\trMmTwentyFour$ and two 2026 models at $\trMmTwentySix$. Agreement with readers rises nearly as fast
(\trMrFactor$\times$ against \trMmFactor$\times$) --- so the familiar claim that capability buys
machine agreement and not human agreement is wrong --- but it arrives at the level two readers reach with each other and stops
there, while agreement between models passes it and keeps climbing, from
\trMmOverHhTwentyFour$\times$ the human yardstick to \trMmOverHhTwentySix$\times$.
\end{abstract}

\section{Introduction}

That language models produce more similar outputs than people do is an increasingly common claim,
supported by studies of writing assistance, idea generation and survey simulation. Almost all of
them share a methodological shape: the human comparison group is recruited, instructed, and paid.
That design answers ``do models differ from instructed humans,'' which is a weaker question than
the one being asked, because instruction is itself a homogenising force. If two crowdworkers given
identical guidelines agree more than two people going about their day, then part of any measured
model--human gap is the guidelines.

This paper measures the same phenomenon against a human reference that was never assembled for a
study. On a social highlighting platform, readers mark passages in web documents for their own
purposes --- to remember, to quote, to return to. They are not told what matters. They are not
compensated. On the platform studied here, the on-page overlay showing other readers' highlights is
off by default and, per the operator, rarely enabled, so readers do not see each other's marks
while reading. Their marks accumulate per document, and on well-read documents a dozen or more
independent readings of the same text exist.

We ask three questions of that substrate.

\begin{enumerate}
\item How much do two language models agree with each other, on a scale set by how much two people
agree, and by how much one person agrees with themselves?
\item Is that a property of language models, or of two frontier systems from the same year prompted
identically?
\item What do models converge \emph{on} that readers do not?
\end{enumerate}

\paragraph{Contributions.}
(i) A convergence measurement whose human baseline is naturalistic and uninstructed, at a scale
lab studies cannot buy. (ii) A position-controlled agreement estimator whose calibration is
\emph{demonstrated} --- the pairs that must be zero are zero --- and whose two prior versions we
report as failures, because the second defect would have manufactured a competing explanation out
of nothing. (iii) A pre-registered \nArms-arm panel, chosen to falsify rather than confirm, that
locates convergence as a graded property of capable models rather than a binary property of
language models. (iv) A self-agreement baseline that reframes the result: a rival lab's model can be
closer to GPT-5.4 than a second call to GPT-4o is to the first.

\section{Related work}

\paragraph{Homogenisation is established; the human baseline is where designs differ.}
Doshi and Hauser~\cite{doshi2024} find that writers given generative-AI ideas produce stories
judged more creative individually and \emph{more similar to each other} collectively. Padmakumar
and He~\cite{padmakumar2024} find that co-writing with a feedback-tuned model reduces content
diversity across authors, and attribute it to the model contributing less diverse text. Both are
controlled experiments in which the human comparison group is recruited and given the same task, so
what they establish is that AI assistance homogenises \emph{instructed} humans. Our contribution is
orthogonal: the same phenomenon measured against people who were given no task at all, which is
the comparison those designs cannot construct.

\paragraph{Instructed humans do not close the gap on their own.}
Gilardi et al.~\cite{gilardi2023} give crowd workers, trained annotators and a language model the
same codebooks on the same texts, and report the model's intercoder agreement above both groups of
humans on average, and above trained annotators on every task at their lower temperature setting.
Reiss~\cite{reiss2023} qualifies that reliability directly, showing the model's annotations are
strongly sensitive to prompt and temperature. Gilardi et al.\ bear on the instruction asymmetry
without removing it: their model figure is one model's consistency across two runs of the same
prompt, while their human figure is between two different people, so the comparison substitutes a
self-versus-cross asymmetry for the instructed-versus-uninstructed one. What it does establish is
that instructing the humans does not by itself close the gap. Their task also has a correct answer
and a codebook that defines it; ours has neither, and its human side was never asked for anything.

\paragraph{Chance-adjusted agreement between models.}
Goel et al.~\cite{goel2025} propose a chance-adjusted probabilistic agreement measure for model
similarity and report that model errors grow more alike as capability rises --- the same class of
instrument, and the same direction, as Section~5 here. What each can claim differs in two ways.
Theirs runs over all predictions on labelled multiple-choice benchmarks, with the chance term
derived from each model's accuracy, so it asks whether models are wrong in the same way; ours runs
over \emph{choices} on a task with no correct answer, so it asks whether they want the same things.
And their chance correction is against per-model accuracy, where ours is against the position and
length of the sentences chosen --- on this substrate the confound that dominates raw overlap
(Section~4). Neither result implies the other; that two different substrates move the same way is
the part worth having.

\paragraph{Monoculture as a systems concern.}
Kleinberg and Raghavan~\cite{kleinberg2021} show that many decision-makers converging on one
algorithm can lower collective decision quality even when that algorithm is individually more
accurate. Bommasani et al.~\cite{bommasani2022} ask whether monoculture produces outcome
homogenisation --- the same people rejected everywhere. Both are arguments about consequences given
convergence. We supply a measurement of convergence itself, on a scale set by human disagreement.

\paragraph{Models as stand-ins for people.}
Argyle et al.~\cite{argyle2023} show conditioned language models can reproduce the response
distributions of human subgroups, which has made model-simulated populations a common research
instrument. Our result bears on that use directly and unfavourably: on this task the models agree
with each other several times more than they agree with readers, and no model in an \nArms-arm
panel agrees with readers more than a reader does. A population simulated from several models is
not several populations.

\paragraph{Extractive salience.}
Selecting important sentences is old~\cite{luhn1958}, and the classical family gives us the control
that matters most here: if any two procedures agreed, our result would be about determinism rather
than about models. They do not (Section~4).

\paragraph{Whether model salience is human salience.}
Trienes et al.~\cite{trienes2025} probe salience behaviourally across thirteen models and four
datasets, tracing which questions a model's summaries keep answerable as the length budget is
squeezed. They report both halves of what we report: that this notion of salience is largely
consistent across model families and sizes, and that it correlates only weakly with human ratings
of question salience. Neither half is ours to claim. What their design does not do --- because their
model--model agreement is a claim-level reliability coefficient and their human--human agreement is
a rank correlation over question-level ratings --- is put those two on one metric, so the sentence
\emph{models agree with each other more than readers agree with each other} is not available to
them. Ours are the same statistic on the same units, and the
human side is what people marked while reading rather than what they said mattered afterwards. Weak
agreement with readers is the half already established; the comparison of the two agreements is
what this paper adds.

That readers agree poorly with each other on which sentences matter is itself old. Rath, Resnick
and Savage~\cite{rath1961} found not only that different subjects select different sentences, but
that the same subjects re-selected differently on a later occasion --- so a low human figure is the
ordinary condition of the task rather than a defect of our readers or our substrate.

\paragraph{The estimator.}
The matched, position-controlled machinery, the corpus and its gates are from our earlier work on
compression fidelity~\cite{paper6}, which measured a single model against crowd highlights. This
paper reuses the substrate to ask a different question --- agreement among predictors rather than
agreement with the crowd --- and reports two corrections to that estimator's null (Section~4). Three
pre-registrations travel with this paper. The panel's was written before any arm in it was called
and fixes what is tested here. The earliest belongs to a different study on the same substrate
--- how large a crowd of readers a model is worth --- whose headline quantity this paper does not
report; it ships because the estimator, corpus gates and kill conditions this paper inherits are
specified in it, not because its hypotheses are ours. The third fixes the reading rule and the stop
threshold for the shared-goal bound reported in Limitations, and it ships because that bound is the
one figure here whose reported form is post-hoc: the reader should be able to check what was fixed
in advance against what was not.

\section{Data}

\paragraph{Documents and readers.}
\corpDocs{} public web documents drawn from \corpDomains{} domains with at most three documents per
domain. Sentences are 30--400 characters; documents are capped at 300 sentences. For each document
we hold uid-free mark index sets for up to 60 readers, shuffled so no reader can be followed across
documents: \textbf{\corpMarkSets{} reader mark sets in total, median \corpMedianReaders{} per
document}, ranging from \corpMinReaders{} to \corpMaxReaders{}. This is the corpus published with
our earlier work on compression fidelity; the gates, extraction and anchoring are unchanged. That
work selected documents with at least twelve highlighters; two documents fall below twelve
\emph{mark sets} here, because a highlighter whose marks all fail to anchor to an extracted sentence
contributes none. We state the realised range rather than the selection gate, since the range is
what the analyses run on.

\paragraph{Independence.}
The design requires that reader $k$ is not copying readers $1..k{-}1$. The platform's on-page
overlay of others' highlights is off by default and rarely enabled. We record this as an
\textbf{assumption}, not a finding: take-up of that setting is not verifiable from the artifacts we
hold. A companion idea --- measuring social contagion in highlighting --- was abandoned on exactly
this fact before it was designed.

\paragraph{Models.}
\nArms{} usable model arms: \nVendors{} vendors (OpenAI, Anthropic, Google, Meta, Mistral, Alibaba,
DeepSeek, Microsoft, IBM, NVIDIA, Z-AI), \nCountries{} countries, \nGenerations{} generations
spanning 2024 to 2026, sizes from 8B to frontier, closed and open weights. Each model receives the
entire document as numbered sentences and returns a ranking of every sentence by importance; the top
$\mathrm{round}(0.2n)$ is its keep set. The prompt is byte-identical to the one used in our earlier
work, and the two incumbent arms reproduce their previously published keep sets in \corpDocs{} of \corpDocs{}
documents.

\section{Measure}

Let $A$ and $B$ be sets of sentences on the same document, each of size $b = \mathrm{round}(0.2n)$.
Readers with at least $b$ marks are truncated uniformly at random to $b$; readers with fewer are
excluded for that document.

\paragraph{Why raw overlap will not do.}
Readers who both mark early, short sentences agree by position alone. Of the \mmRaw\% raw overlap
between our two incumbent models, \posShare{} points are accounted for by position. Every figure we
report is therefore \emph{excess} agreement:
\[
\mathrm{excess}(A,B) \;=\; \frac{|A \cap B| - \mathbb{E}\,|A' \cap B'|}{b},
\]
where $A'$ and $B'$ are independent resamples of $A$ and $B$ within depth-and-length bands: sentence
$y$ may replace $x$ if it lies within $0.10$ relative depth and $0.10$ within-document length rank
of $x$. On the median document a band holds \bandMidMedian{} sentences and
\bandMidSingleton\% of sentences have no permissible replacement but themselves, which is why the
tolerance is $0.10$ and not the $0.05$ our earlier work used: at $0.05$ that share is
\bandLowSingleton\% and the null is partly degenerate (Section~5.7).

\paragraph{The null must preserve set size, and ours did not, twice.}
Two earlier versions of this null are reported here as failures rather than omitted.

\emph{First}, the expectation was computed as a sum of per-element weights where it needed inclusion
probabilities. Against a \rSevenSims-draw simulation the sum form erred by \rSevenSumModel{}
sentences on the model pair and \rSevenSumHuman{} on a reader pair; the inclusion form errs by
\rSevenInclModel{} and \rSevenInclHuman{}. Because the expectation is subtracted, this understated
all excess figures, and understated positionally concentrated arms most.

\emph{Second}, and worse, resampling each element independently let the set \emph{collapse} when two
elements landed on the same sentence --- unequally, retaining \shrinkLead{} of the budget for naive
truncation, \shrinkGpt{} and \shrinkOpus{} for the models, \shrinkReader{} for readers and
\shrinkRandom{} for a random baseline. Expected overlap was thus computed on shrunken sets while
observed overlap used full ones, paying a bonus for clustering, which is the exact artifact this
null exists to remove. \textbf{The tell was that naive truncation scored well above zero against a
random baseline, where two independent selectors must score zero.} The null in use draws without
replacement and preserves set size exactly, which an assertion enforces; inclusion probabilities are
estimated by simulation once per set and reused across pairs.

\paragraph{Calibration, demonstrated.}
Every one of the \randN{} pairs involving the random baseline lands within \randMaxAbs{} of zero,
and all \clasN{} classical--classical pairs lie between $\clasMin$ and $\clasMax$. Under the
pre-correction null the same table placed two classical pairs above the human yardstick --- which
would have made ``any two algorithms agree'' a live competing explanation built entirely from
estimator error.

\paragraph{Which documents each analysis uses, and why the human yardstick moves.}
Different questions need different minimum reader counts, so the analyses run on nested subsets of
the \corpDocs{} documents, and the same quantity therefore appears with slightly different values in
different sections. The panel needs two size-matched readers per document (\nDocsPanel{}
documents, human--human $\hhEst$). The ceiling analysis needs three, because a leave-one-out
consensus of the others requires at least two to build it (\ceilDocs{} documents, human--human
$\ceilTwoReadersEst$), and the saturation curve needs nine (\ceilCurveDocs{} documents). The
characterisation needs both consensus sets to be non-empty and disjoint (\charDocsN{} documents),
and the reader-equivalence curve needs sixteen readers (\curveDocsN{} documents). \textbf{Every
comparison is made within one subset; nothing is compared across them.}

\paragraph{Inference.}
Domain-clustered bootstrap, 10{,}000 resamples, percentile intervals. Every model-versus-human
contrast is paired per document before bootstrapping.

\section{Results}

\subsection{One scale}

\begin{table}[h]
\centering
\begin{tabular}{lr}
\toprule
comparison & excess agreement \\
\midrule
GPT-5.4 vs \textbf{itself} (same prompt, fresh call) & $\selfGptEst$ [$\selfGptLo$, $\selfGptHi$] \\
GPT-5.4 vs \textbf{itself}, paraphrased prompt & $\paraGptEst$ [$\paraGptLo$, $\paraGptHi$] \\
a reader vs \textbf{themselves} (second truncation) & $\ceilingEst$ [$\ceilingLo$, $\ceilingHi$] \\
\textbf{GPT-5.4 vs Claude Opus 5} (rival labs) & $\mathbf{\mmEst}$ [$\mmLo$, $\mmHi$] \\
the best-agreeing reader pair per document & $\hhBestEst$ [$\hhBestLo$, $\hhBestHi$] \\
Llama 3.1 70B vs \textbf{itself}, paraphrased & $\paraLlamaEst$ [$\paraLlamaLo$, $\paraLlamaHi$] \\
GPT-4o direct vs GPT-4o via a proxy & $\routeFouroEst$ [$\routeFouroLo$, $\routeFouroHi$] \\
GPT-4o vs \textbf{itself} (same provider, second call) & $\selfFouroEst$ [$\selfFouroLo$, $\selfFouroHi$] \\
\textbf{two different human readers} & $\mathbf{\hhEst}$ [$\hhLo$, $\hhHi$] \\
\bottomrule
\end{tabular}
\caption{All quantities on one scale, \nDocsPanel{} documents.}
\label{tab:scale}
\end{table}

\paragraph{One pre-registered kill condition fired, and the diagnosis is a result.} The panel
pre-registration required self-agreement to exceed every cross-model pair, on the reasoning that if
it did not, something was wrong with the comparison itself. One of \nPairs{} pairs
exceeds it: \killPairA{} against \killPairB{} reaches $\killPairEst$ [$\killPairLo$,
$\killPairHi$] where GPT-5.4 against a fresh call of itself reaches $\selfGptEst$. No pair clears
the self-agreement \emph{interval}, so the two are not separated by our sampling. We
report it because the diagnosis is not an error: sampling makes self-agreement less than unity,
and two models from one vendor and one generation can be as alike as one model is to its own second
answer. That is the paper's thesis in its sharpest available form, and we would have missed it by
treating a fired condition as a fault.

\textbf{A reader shares \readerCeilShare\% of their own self-agreement with another person; the two
models reach \modelCeilShare\% of it with each other.} (``Ceiling'' is reserved below for a
different and more useful quantity --- what a predictor achieves against the crowd, measured in
Section~5.4 --- so it is not used for self-agreement here.) Put the other way: two frontier models
from rival labs agree \mmOverSelfFouro{} times as much as GPT-4o agrees with itself.

\subsection{What the scale means}

Excess agreement counts sentences. Each party names $\mathrm{round}(0.2n)$ of them, so multiplying
by that budget converts any figure here into sentences shared beyond what position and length
already predict. On the median document --- \intDocSents{} sentences, of which each party names
\intBudget{}:

\begin{center}
\begin{tabular}{lrrrr}
\toprule
 & observed & chance & excess & of \intBudget{} named \\
\midrule
two readers & \intHhObs\% & \intHhChance\% & \intHhExtra{} sentences & \intHhShared{} shared \\
two models & \intMmObs\% & \intMmChance\% & \intMmExtra{} sentences & \intMmShared{} shared \\
\bottomrule
\end{tabular}
\end{center}

So of \intBudget{} sentences each names, two readers share about \intHhShared{} and two models
about \intMmShared{}; and once position is accounted for, the readers' shared choices amount to
\intHhExtra{} sentences and the models' to \intMmExtra{}. In the same units the crowd-consensus
ceiling is worth \intCeilSents{} sentences and the best arm reaches \intBestSents{}. \textbf{Two
models share \intModelOverCeil{} times as many extra sentences with each other as perfect knowledge
of the crowd buys with a reader.}

\subsection{The panel}

Across \nPairs{} model pairs the median is $\pairMedian$ (quartiles $\pairPtwentyfive$ and
$\pairPseventyfive$) against a human yardstick of $\hhEst$; \textbf{\nAboveHuman{} pairs sit
entirely above the human interval and \nPairsBelowHuman{} fall below its point estimate}. Our
pre-registration fixed the \emph{mean} over pairs as the pooled quantity and we report the median,
because one pair at $\pairMax$ moves a mean of \nPairs{}. Both are given so the substitution is
visible rather than trusted: the mean is $\pairMean$, \pairMeanOverHh$\times$ the human yardstick,
against \pairMedianOverHh$\times$ for the median.

\textbf{The pair we quote most is not a typical pair.} GPT-5.4 against Claude Opus 5 is the \headlinePctile{} percentile of the panel. The \nCrossPairs{} cross-vendor pairs have a median of
$\crossMedian$, \crossMedianOverHh$\times$ the human figure rather than \mmOverHh$\times$. Both
belong in the paper: the incumbent pair is what our earlier work measured and is the sharpest
illustration, and the cross-vendor median is what a reader should carry away as typical.

Convergence is graded, and the grading is the result (Table~\ref{tab:groups}).

\begin{table}[h]
\centering
\begin{tabular}{lrrl}
\toprule
grouping & pairs & median & range \\
\midrule
both closed & \grpClosedN & $\grpClosedMed$ & $\grpClosedMin$ to $\grpClosedMax$ \\
both frontier & \grpFrontierN & $\grpFrontierMed$ & $\grpFrontierMin$ to $\grpFrontierMax$ \\
cross-everything & \grpCrossN & $\grpCrossMed$ & $\grpCrossMin$ to $\grpCrossMax$ \\
different vendor & \grpDiffVendorN & $\grpDiffVendorMed$ & $\grpDiffVendorMin$ to $\grpDiffVendorMax$ \\
both open & \grpOpenN & $\grpOpenMed$ & $\grpOpenMin$ to $\grpOpenMax$ \\
\textbf{both small} & \grpSmallN & $\mathbf{\grpSmallMed}$ & $\grpSmallMin$ to $\grpSmallMax$ \\
\midrule
same country & \ctrySameN & $\ctrySameMed$ & --- \\
different country & \ctryDiffN & $\ctryDiffMed$ & --- \\
\midrule
\emph{two human readers} & --- & \emph{$\hhEst$} & --- \\
\bottomrule
\end{tabular}
\caption{Pairs sharing nothing --- different vendor, country, size tier and weight regime --- still
reach $\grpCrossMed$, \crossOverHh$\times$ the human yardstick. The smallest models reach the human
level and no more. Pairs spanning a national border agree \emph{more} than pairs inside one
($\ctryDiffMed$ against $\ctrySameMed$), which is the wrong direction for a shared-corpus or
national-style explanation.}
\label{tab:groups}
\end{table}

\subsection{Both kinds of agreement are rising. Only one has room left}

Agreement with human readers runs from $\typMin$ (\typMinName) to $\typMax$ (\typMaxName),
against a reader's $\hhEst$ with other readers. \nTypAboveHuman{} models exceed that point
estimate and none exceeds it beyond sampling noise: no interval clears the human figure.

An earlier version of this paper concluded from that ``capability buys agreement with other models,
not with readers.'' \textbf{Measuring it showed the claim is false}, and the correction is the more
interesting result.

Pair agreement is monotone in both scale and recency. By size tier it runs from $\tierSmallSmall$
for two small models to $\tierFrontFront$ for two frontier ones, through every intermediate cell;
by generation from $\trMmTwentyFour$ for two 2024 models to $\trMmTwentySix$ for two 2026 models.

\textbf{We cannot separate the two, and the panel is why.} It was built to \emph{span} vendor,
country, generation, size and weight regime, not to \emph{balance} them, and generation came out
badly confounded with the rest: our 2024 arms are one closed and four open with no frontier model,
our 2026 arms are five closed and one open with three. So ``newer models agree more'' and ``closed
frontier models agree more'' are the same statement in this design, and the marginal correlations
($\genCorr$ for generation, $\tierCorr$ for tier) cannot be read as separating them. Inside a
stratum the cells move the same way --- among open models, $\wsOpenEarly$ in \wsOpenEarlyYear{} against $\wsOpenLate$ in
\wsOpenLateYear{}; among frontier models, $\wsFrontEarly$ in \wsFrontEarlyYear{} against
$\wsFrontLate$ in \wsFrontLateYear{} --- but those cells hold between \wsCellMin{} and
\wsCellMax{} pairs. \textbf{We report the trend as suggestive and unseparated.} A panel
balanced on generation would settle it and ours is not one.

But agreement with \emph{readers} rises with generation too, and at a similar factor:

\begin{center}
\begin{tabular}{lrrr}
\toprule
 & 2024 models & 2026 models & factor \\
\midrule
agree with each other & $\trMmTwentyFour$ & $\trMmTwentySix$ & \trMmFactor$\times$ \\
agree with readers & $\trMrTwentyFour$ & $\trMrTwentySix$ & \trMrFactor$\times$ \\
\midrule
\emph{as a multiple of two readers} & \trMmOverHhTwentyFour$\times$ & \trMmOverHhTwentySix$\times$ & \\
\bottomrule
\end{tabular}
\end{center}

\textbf{Newer models read more like a person \emph{and} much more like each other, at
comparable rates. What differs is how much room each has left}, and establishing that required
measuring a ceiling rather than assuming one.

An earlier version of this section argued that a model cannot agree with readers more than readers
agree among themselves. That is false: a predictor aimed at the \emph{centroid} of a crowd beats
any two members whenever they share signal and differ in noise. On the \ceilDocs{} documents with
at least three size-matched readers, the leave-one-out crowd consensus --- the other readers'
majority, scored against the held-out reader --- reaches $\ceilCentroidEst$
[$\ceilCentroidLo$, $\ceilCentroidHi$] where two individual readers reach $\ceilTwoReadersEst$.
The pairwise figure is \emph{not} a bound; it is \ceilOverPairwise$\times$ below the consensus figure.

All figures in this subsection are computed on the \ceilDocs{} documents with at least three
size-matched readers, where two readers reach $\ceilTwoReadersEst$ and two models
$\ceilTwoModelsEst$ --- the same quantities as elsewhere in the paper, on the smaller sample this
analysis requires.

The ceiling itself depends on how many readers build the consensus, so a consensus of three is not
the ceiling unless the curve has flattened. On the \ceilCurveDocs{} documents carrying enough
readers to trace it, it rises from $\ceilCurveFirst$ at \ceilCurveFirstK{} readers and flattens near $\ceilSat$ by
\ceilSatK{}. We take $\ceilSat$ as the ceiling and note three things about it: it rests on
\ceilCurveDocs{} documents with wide intervals; it is a maximum over the five consensus sizes we
traced, which is the same selection-over-noise we flag below for ``the best of \nArms{} arms'';
and that selection is worth \ceilSpanTopTwo{} here, the gap to the next-largest point. Taking the
maximum biases the ceiling \emph{up}, which shrinks every share and every multiple computed
against it, so each claim below is the conservative end.

Against it:

\begin{center}
\begin{tabular}{lrr}
\toprule
 & excess agreement & share of the ceiling \\
\midrule
crowd consensus of six readers vs a held-out reader & $\ceilSat$ & 100\% \\
\armMaxName{} --- \emph{best of \nArms{} arms} & $\armMaxEst$ & \armMaxShare\% \\
GPT-5.4 --- pre-specified incumbent & $\armIncumbentEst$ & \armIncumbentShare\% \\
\armMedianName{} --- median arm & $\armMedianEst$ & \armMedianShare\% \\
two individual readers & $\ceilTwoReadersEst$ & --- \\
\midrule
\textbf{two models} & $\mathbf{\ceilTwoModelsEst}$ & \textbf{\ceilModelOver$\times$} \\
\bottomrule
\end{tabular}
\end{center}

\textbf{Saturation is a property of the best arms, not of models.} The top arm has taken
\armMaxShare\% of what is achievable on the reader side --- though that figure is a maximum over
\nArms{} and biased upward --- the pre-specified incumbent \armIncumbentShare\%, and the median
arm \armMedianShare\%. Meanwhile two models stand at \ceilModelOver{} times the same ceiling with
each other. The best models are running out of room on the reader side; typical ones are not; and
nothing is running out of room on the machine side.

\subsection{What they converge on: not register}

If capable models pick sentences readers do not, the cheapest explanation is that they prefer a
different \emph{kind} of sentence --- the encyclopedic register, say. We tested that on
\chrNFeatures{} surface features over \chrDocs{} documents and all \chrArms{} usable arms,
comparing the sentences only the models chose against those only the readers chose.

Two earlier versions of this comparison each produced a different answer, and both are reported
because the difference between them is the methodological point.

\emph{Without a length control} it appeared to confirm the register story, with definitional verbs,
capitalised words, first person and multi-clause sentences all separating the groups. It was a
length artifact: the sentences only models chose are \chrLenGap{} characters longer
[\chrLenGapLo, \chrLenGapHi] and \chrDepthGap{} shallower in relative depth than the ones only
readers chose, and nearly every feature here rises with sentence length, because
a longer sentence has more chances to contain a comma, a capitalised word or a copula.

\emph{With a hand-picked arm set} --- eight arms ``chosen as the ones the panel showed converge'' ---
stratifying by depth and length terciles left nothing significant. But that set was selected after
seeing which arms converge, and it mattered: under every usable arm, two features become nominally
significant. A characterisation whose conclusion moves with a post-hoc choice of arms is reporting
the choice, so the arm set is now the coverage rule the pre-registration already fixed.

\paragraph{The result.} Stratified within document depth and length terciles, over all usable arms,
two of \chrNFeatures{} features clear zero at a nominal 95\%: definitional verbs at
$\chrDefDiff$ [$\chrDefLo$, $\chrDefHi$] and connective openings at $\chrConnDiff$, whose
lower bound is $\chrConnLo$. \textbf{Neither survives a Bonferroni correction over the
\chrNFeatures{} features tested; nothing does.}

\textbf{So the divergence is not register.} At equal depth and equal length, models and readers
choose sentences with the same surface properties; they simply choose different sentences. Whatever
separates them is not visible in the vocabulary, punctuation or person of the sentence. We report
that rather than a table that survives only without a control the rest of this paper insists on,
or only under an arm set we chose after looking.

\subsection{Out of sample: models released after the analysis}

Everything above was fixed before these arms existed, which makes them the cleanest test this paper
can be given. The claim under test is not that models agree --- it is that \textbf{agreement
between models keeps climbing while agreement with readers arrives at roughly the level two readers
reach and stops}. A model released after the analysis either continues that or breaks it.

Checking the provider's model list returned no plain successor but three siblings, plus one
intermediate release the panel does not contain: \osNArms{} arms in all.
\textbf{They are reported separately and enter no pooled statistic in this paper} --- not the
\nPairs{} pairs, not a group median, not the generation trend --- which is what our
pre-registration requires of anything added after results are seen. Predictions were written and
committed before the first call (\texttt{PREREG-PANEL.md}, DEV-4). Of \osNPred{}, \osHeld{} held and \osFailed{} failed.

\paragraph{The measurement is the same measurement.} Every arm the two analyses share reproduces
the panel to within \osScaleWorst{} --- Monte Carlo noise between two independent draws of the
null. Without that, nothing below would be comparable to anything above.

\paragraph{Between models: as predicted.} Against the panel's three frontier incumbents the
\osPairN{} new pairs have a median of $\osPairMedian$, \emph{above} the incumbent pair this paper
quotes most. None of the new arms returned a degenerate ranking.

\paragraph{With readers: the ceiling holds, and it holds at a stronger test than we set.}

\begin{center}
\begin{tabular}{lrl}
\toprule
arm & agreement with readers & \\
\midrule
\osFiveName & $\osFiveEst$ [$\osFiveLo$, $\osFiveHi$] & \\
\osLunaName & $\osLunaEst$ [$\osLunaLo$, $\osLunaHi$] & \\
\osSolName & $\osSolEst$ [$\osSolLo$, $\osSolHi$] & \\
\osTerraName & $\osTerraEst$ [$\osTerraLo$, $\osTerraHi$] & \\
\midrule
\emph{two human readers} & \emph{$\hhEst$ [$\hhLo$, $\hhHi$]} & \\
\emph{best of the \nArms{} panel arms} & \emph{$\armMaxEst$} & \\
\emph{crowd-consensus ceiling} & \emph{$\ceilSat$} & \\
\bottomrule
\end{tabular}
\end{center}

\osAbovePoint{} of \osNArms{} sit above the human point estimate against
\osPanelAbovePoint{} of \nArms{} in the panel --- a larger share, though on \osNArms{} arms that
is one arm's worth of difference and we do not read anything into it. \textbf{None clears the
human interval}, which is the test that would falsify the paper, and none clears it after a
Bonferroni widening over all \osMultN{} new intervals either. No new arm reaches the best panel
arm, and none approaches the ceiling.

\paragraph{What we got wrong, and the measurement it forced.} We predicted the three siblings
would agree with each other at $\osSibPredicted$ or above, near the level a model reaches with its
own second answer, on the reasoning that same-vendor pairs are the panel's highest. They do not:
their median is $\osSibMedian$ and their maximum $\osSibMax$.

A raw pair value cannot carry that comparison on its own, because self-agreement is model-specific
in this panel --- GPT-5.4 reaches $\selfGptEst$ with a fresh call of itself and GPT-4o only
$\selfFouroEst$ --- so we called each sibling a second time under the byte-identical prompt and
measured what it reaches against \emph{itself}. Each exceeds what it reaches with its siblings
($\osSelfMin$ to $\osSelfMax$, mean $\osSelfMean$), so this is not a second instance of the kill
condition above. And it gives the comparison the scale it was missing:

\begin{center}
\begin{tabular}{lrr}
\toprule
 & agreement & as a share of self-agreement \\
\midrule
three variants of \emph{one} vendor & $\osSibMedian$ & \osWithinRetain\% \\
two models from \emph{rival} laboratories & $\mmEst$ & \osCrossRetain\% \\
\bottomrule
\end{tabular}
\end{center}

\textbf{Measured against what each retains of its own self-agreement, a trio of variants from one
vendor and a pair from rival laboratories come out close.} We say close and not equal: the two
shares are not built from identical quantities --- the first divides a median over three sibling
pairs by a mean over three self-agreement arms, the second divides one pair by one model's own
figure --- and neither ratio carries an interval. That is a failed prediction, and it strengthens
rather than weakens the paper's noun: convergence is not a within-vendor artifact, because
within-vendor turns out not to be special. The second failed
prediction is the ladder below.

\paragraph{A recency ladder the panel could not supply, and what it cannot settle.} The paper
reports its generation trend as suggestive because recency is confounded with weight regime and
scale. These arms give a ladder with vendor, weight regime and size tier \emph{fixed}. Along it,
agreement with readers stays between $\osLadderReaderMin$ and $\osLadderReaderMax$ and agreement
with the panel's incumbent between $\osLadderMachineMin$ and $\osLadderMachineMax$, and
\textbf{neither is ordered by recency} --- the machine side is not monotone along the ladder, so
its spread is noise around a level rather than a climb. \emph{We predicted a rise here and did not
get one.} We are careful about what that licenses. The panel's trend spans two years and this
ladder spans a few months, so a flat ladder is \emph{not} evidence against a trend over years. What
it shows is that within one vendor over one short interval, recency alone moved neither side --- an
additional reason to read the trend as suggestive, not a refutation of it.

\subsection{Sensitivity}

\paragraph{The band width.} It is the null's one free parameter. Widening it moves the magnitudes
and not the ordering: at a tolerance of $0.05$ the human and model figures are $\swpLowHh$ and
$\swpLowMm$ (difference $\swpLowDiff$); at $0.20$ they are $\swpHighHh$ and $\swpHighMm$
(difference $\swpHighDiff$). \textbf{The low arm is not a stricter null but a partly degenerate
one}, and this is the reason we do not run the estimator there. At $0.05$ the median band holds
\bandLowMedian{} sentences and \bandLowSingleton\% of sentences have no permissible replacement
but themselves; a sentence that cannot be moved contributes equally to observed and expected
overlap and nets to nothing, so a quarter of the corpus is inert and the collapse of the magnitudes
is partly arithmetic. At our $0.10$ the median band holds \bandMidMedian{} and singletons are
\bandMidSingleton\%; at $0.20$, \bandHighMedian{} and \bandHighSingleton\%. The ordering
survives all three, which is the claim; the magnitudes should be read only at $0.10$ and above.

\paragraph{The two random number generators underneath.} The null's inclusion probabilities are
estimated by simulation, and readers with more than $b$ marks are truncated to $b$ at random. Both
were fixed once and never varied, so we varied them. Across simulation budgets from \mcLowSims{}
to \mcHighSims{} the gap moves by \mcShiftShare\% of its own interval --- the Monte Carlo budget
is not a live parameter. The truncation draw matters more and still not much: over \seedN{}
independent draws the human figure ranges $\seedHhMin$ to $\seedHhMax$ and the gap
$\seedGapMin$ to $\seedGapMax$, a span of \seedSpanShare\% of the reported interval, with every
draw above zero. \textbf{That variance is not inside our intervals}, which resample documents and
hold the truncation fixed; it is small enough not to change any statement here, and we report it
rather than let a reader assume it was included.

\section{Robustness}

Four threats had never been examined and were taken in a zero-based pass; four more test whether
the result depends on our own choices rather than on the data.

\paragraph{Document drift is the one asymmetry we cannot design away, and it does not bite.}
Readers marked these documents over roughly seven years; every model read one 2026 snapshot. A
document that changed in between leaves a reader's marks anchored to text they never saw, and the
models never pay that cost --- which is exactly the direction that would inflate our headline. Per
document we hold a mark-month histogram, so the test is differential: as a document's marks spread
over more time, human agreement should fall if drift matters and model agreement should not move.
Splitting at a median span of \driftCut{} months (range up to \driftSpanMax{}), human agreement is
$\driftNarrowHh$ on narrow-span documents and $\driftWideHh$ on wide-span ones --- \emph{higher}
on the wide ones --- while model agreement moves the same way ($\driftNarrowMm$ to
$\driftWideMm$). Both arms shift together, so this is a property of those documents rather than of
drift, and human agreement is not selectively depressed.

\paragraph{Not a granularity artifact.} Excess is divided by a budget of $\mathrm{round}(0.2n)$,
so short documents give coarse values. The gap is $\sizeShortGap$ on the shorter half and
$\sizeLongGap$ on the longer half, with both lower bounds above zero ($\sizeShortLo$,
$\sizeLongLo$).

\paragraph{No single domain carries it.} The \nDocsPanel{} documents that carry at least two
size-matched readers span \lodoNDomains{} of the corpus's \corpDomains{} domains. Leaving out each of those
\lodoNDomains{} in turn moves the paired gap by at most \lodoWorst.

\paragraph{Multiplicity.} We report \nPairs{} pair intervals, of which \multRaw{} lie entirely above
the human interval at 95\%; after a Bonferroni widening to simultaneous coverage, \multBonf{} still do.

\paragraph{The models are cut sharply and the readers are not, and that is worth half the gap.}
This is the asymmetry we take most seriously, and it had to be measured rather than argued. A
model's set is its \emph{top} $b$. A reader's is a uniformly random $b$ of the marks they made,
because a reader supplies no ranking and there is no principled way to take their best $b$. The
median reader marks \symOvermarkMedian$\times$ the budget, so a third of what they said is
discarded by a coin flip and the models never pay that.

So we put the models through the readers' procedure exactly: for each reader holding $M$ marks,
each model is cut to its top-$M$ --- the same amount that reader said --- and then a uniformly
random $b$ is drawn from it. \textbf{Blunting the models the way readers are blunted costs them
\symLoss\% of their agreement}, from $\symSharpEst$ to
$\symMatchedEst$ [$\symMatchedLo$, $\symMatchedHi$], against readers at $\symHhEst$. The paired
gap falls from $\symGapSharpEst$ to $\symGapMatchedEst$ [$\symGapMatchedLo$,
$\symGapMatchedHi$] --- \symMatchedOverHh$\times$ the human figure rather than
\symSharpOverHh$\times$, still entirely above zero.

Because both sides shed agreement as more is discarded, the fair test is to compare them at equal
discarding. Splitting reader pairs by how much they over-marked and blunting the models to each
level --- keeping only the \homoTotal{} pairs whose two readers fall in the same band, for the
reason given below:

\begin{center}
\begin{tabular}{lrrrrrl}
\toprule
over-marking & pairs & docs & domains & two readers & two models & gap \\
\midrule
\homoALabel & \homoAN & \homoADocs & \homoADoms & $\homoAHh$ & $\homoAMm$ & $\homoAGap$ [$\homoALo$, $\homoAHi$] \\
\homoBLabel & \homoBN & \homoBDocs & \homoBDoms & $\homoBHh$ & $\homoBMm$ & $\homoBGap$ [$\homoBLo$, $\homoBHi$] \\
\homoCLabel & \homoCN & \homoCDocs & \homoCDoms & $\homoCHh$ & $\homoCMm$ & $\homoCGap$ [$\homoCLo$, $\homoCHi$] \\
\homoDLabel & \homoDN & \homoDDocs & \homoDDoms & $\homoDHh$ & $\homoDMm$ & $\homoDGap$ [$\homoDLo$, $\homoDHi$] \\
\bottomrule
\end{tabular}
\end{center}

Three conventions in that table are worth stating, because each was wrong in an earlier version and
each moved a row. \emph{First}, only pairs whose \emph{both} readers fall inside the band are
counted. Bucketing a pair by the average of its two readers puts a pair with one light and one
very heavy marker into the heavy band while one of its models is barely blunted, and that alone was
worth the bottom row's sign. \emph{Second}, each model is blunted to its own counterpart reader's mark
count, not both to one level, so the two sides are asymmetric in the same way; which model is
paired with which reader does not matter ($\symSwapEst$ [$\symSwapLo$, $\symSwapHi$] between the
two assignments). \emph{Third}, documents and domains are shown because pairs are not the
effective sample --- pairs on one document share readers and the bootstrap resamples domains.

\textbf{The gap holds at every level where the measurement has power, and the bottom row has
none.} The ordering is clear on the first three bands and their intervals exclude zero.
On the fourth --- pairs where \emph{both} readers marked at least $2.5\times$ the budget --- two
readers reach $\homoDHh$ and two models blunted the same way reach $\homoDMm$: \textbf{both are
at chance, and the gap is $\homoDGap$ [$\homoDLo$, $\homoDHi$], which contains zero on
\homoDDoms{} domains}. We read that as a fact about the measurement rather than about models.
Sampling $b$ sentences from a pool of more than $2.5b$ dilutes any set enough that nothing agrees
beyond chance, models included. The comparison stops being informative there; it does not reverse.

An earlier version of this paragraph, written one audit round before this one, claimed the gap was
above zero at \emph{every} level. It was, under a bucketing that mixed a lightly-blunted model
into the heavy band. It is not, done properly.

Two conclusions, and we separate them deliberately. \textbf{The ordering is robust to the
procedure and the magnitudes are not.} Every multiple quoted in this paper is a property of the
comparison as run --- each party's own output, cut as that party produced it --- and a defensible
alternative halves it. We report the published figures as the primary ones, because they compare
what each party actually said, and we state here that they are not procedure-invariant, which no
earlier version of this paper did.

\paragraph{The ordering does not need the bootstrap.} A paired sign test over documents, which
assumes only exchangeability, puts model--model above human--human on \signHigher{} of
\signDocs{} documents (exact two-sided $p = \signP$).

\paragraph{Nor the normalisation.} Dividing excess by the union of the two sets rather than by the
budget gives $\unionGapEst$ [$\unionGapLo$, $\unionGapHi$].

\paragraph{Reader selection makes our figure conservative, not generous.} Only readers with at
least $b$ marks enter, which selects toward heavy markers, and this is the threat we took most
seriously. Lowering the budget ratio admits more readers and re-cuts the model sets from the same
cached rankings, so both sides move together. At a ratio of $0.10$ --- \selLowReaders{} readers per
document instead of the paper's --- human agreement is $\selLowHh$, model agreement $\selLowMm$
and the gap \emph{widens} to $\selLowGap$. At $0.30$, with the most heavily selected readers, the
gap narrows to $\selHighGap$. \textbf{The less selected the readers, the larger the gap} --- but
\ratAsymShare\% of that widening is the asymmetry of the previous paragraph, not the effect, and
we had read the two as independent reassurance. Halving the budget does two things at once: it cuts
each model to a \emph{sharper} set, whose agreement rises from $\ratRefModelSharp$ to
$\ratLowModelSharp$ because the top tenth of a document is more concentrated than the top fifth,
while leaving each reader discarding twice as much. Blunting the models at each ratio separates the
two. From $0.20$ to $0.10$ the published gap widens by \ratSharpWiden{} and the blunted gap by
\ratBluntWiden{}. What survives is real and about a third of what the uncorrected arm suggests:
loosening selection does move the gap our way, and by much less than it appeared to.

\section{What kills which explanation}

Every row but two was pre-registered as a possible outcome rather than constructed afterwards.
The first exception is the shared goal, which comes from a later pilot on a subset of these
documents and whose reported form is post-hoc; it is discussed in Limitations. The second is routing, which was added as a deviation (DEV-2) once we saw that our open-weight
arms reach us through a proxy and our closed ones do not, so openness and routing would otherwise
have been the same variable. It is marked in the table.

\begin{table}[h]
\centering
\begin{tabular}{ll}
\toprule
competing explanation & evidence against \\
\midrule
models are deterministic & GPT-5.4 vs itself is $\selfGptEst$, not unity \\
the shared prompt's wording & paraphrase retains \paraRetain\% of self-agreement \\
the shared prompt's \emph{goal} \emph{(added, post-hoc)} & a different reading goal retains \goalRetain\% \\
any two procedures agree & \clasN{} classical pairs, median $\clasMed$ \\
one vendor's house style & highest cross-vendor pair is $\topCrossEst$ \\
capability converging on truth & no arm's interval clears the human figure \\
proxy routing artifacts \emph{(added, DEV-2)} & paired direct-vs-proxied contrast $\routingEst$ \\
two models is not a panel & \nArms{} arms, \nPairs{} pairs, \nAboveHuman{} above the human interval \\
\bottomrule
\end{tabular}
\caption{Falsifiers and their outcomes. Every row but two was pre-registered; the two exceptions are marked.}
\end{table}

\section{Limitations}

\paragraph{Models were given a task; readers were not.} This is the one confound the design cannot
remove, and it belongs here and in the abstract rather than in a footnote. There is no way to hand
a reader the prompt, so the contribution of \emph{having} an instruction is not measurable on this
substrate. Between that and the wording of the instruction, however, sits a third thing that is:
the \emph{goal} the instruction names. We bound it.

Both quantities below are agreement with a second, independent answer from the same model, divided
by that model's agreement with itself under the same instruction, because the statistic's ceiling is
arm-dependent and raw cells are not comparable across arms. On \goalDocs{} documents, changing the
wording while keeping the goal retains \goalParaRetain\% of self-agreement; changing the \emph{goal}
--- from compressing by importance to predicting what a reader would mark --- retains
\goalRetain\%, an excess of $\goalCrossEst$ [$\goalCrossLo$, $\goalCrossHi$] against a
self-agreement of $\goalSelfEst$. Against that, two readers reach $\goalHumanHumanEst$ where one
reader against a second truncation of their own marks reaches $\goalHumanSelfEst$ ---
\goalReaderShare\%. What the goal change fails to do is close the gap: it still leaves a model
handed a different goal more than twice as self-agreeing as two people are with each other. The
pre-registration is explicit that a low figure here licenses no inference in the other direction,
and we draw none.

Four things bound how far that can be taken, and they are the reason this is a limitation rather
than a result. \emph{First, it is not separable from degradation.} The reader-purpose arm agrees
with actual readers \emph{less} than the incumbent does ($\goalReadVsReadersEst$ against
$\goalCompVsReadersEst$), so part of the drop is a weaker selector rather than a reoriented one, and
nothing here separates the two. \emph{Second, the reported statistic is post-hoc.} The
pre-registration fixed the raw figure and a threshold before the number existed and ships with this
paper; the rescaling to each arm's ceiling was added after that number was seen, and on the
pre-registered raw scale the stop rule did not fire. \emph{Third, the human denominator is not the
model denominator}: a model against a fresh call has generative variability, a reader against a
second truncation of their own marks has none, so that figure measures over-marking and is an upper
bound on the human ceiling --- which makes \goalReaderShare\% an under-statement of the human
share, and the factor above an over-statement, in our own favour.
\emph{Fourth, the subset is not the corpus.} \goalShortDropped{} of \corpDocs{} documents are
dropped because the reader-purpose arm named fewer sentences than the budget and \goalFewReaders{}
more for having fewer than two size-matched readers. The documents dropped for a short answer are
systematically the long ones (median \goalDroppedMedianN{} sentences against \goalMedianN{} for
those kept), so this is measured on the shorter documents. Section~6 reports the model--human gap
itself as smaller on short documents than on long ones, but that is a different quantity on a
different subset, and the direction in which this restriction cuts is not established.

Of the three differences separating the two arms, the goal and the system message push agreement
down; the output protocol's sign is not established, and on the documents kept the
select-at-most-$k$ arm returned exactly the budget, which is the regime in which it most resembles a
truncated ranking. The defensible claim is therefore narrow and unchanged: models under a shared
task converge far more than people reading naturally do, and giving one model two different goals
does not close the gap.

\paragraph{No model with published training data.} The panel was designed to include OLMo-3,
whose training corpus is public, precisely to test whether convergence survives a disjoint corpus.
It has no serving endpoint and no substitute exists. Every remaining arm is trained on undisclosed
data, so we can show convergence survives different vendors, countries, generations, sizes and
weight regimes, and we cannot show it survives a different corpus --- nor a different architecture,
since the panel's only non-transformer fell below the coverage floor.

\paragraph{Anchoring is one-sided.} Reader marks are anchored from highlighted text to sentence
indices; model keep sets are sentence indices directly. Anchoring noise would depress the human
side only, inflating the gap. It is bounded but not eliminated. Under this paper's corrected null,
the consensus of the other readers predicts a held-out reader at $\ceilCentroidEst$
[$\ceilCentroidLo$, $\ceilCentroidHi$] on \ceilDocs{} documents, which mostly-noise marks could not
do. Our earlier work put a single anchored reader against the crowd at $\singleReaderPaperSix$;
\textbf{that figure is computed under \emph{that} paper's null}, the pre-correction one whose two
defects are documented in Section~4, and we label it rather than quoting it as current. Both
defects understate excess, so it is a lower bound on what the corrected estimator would give.

\paragraph{We cannot count distinct readers, or check whether they overlap across documents.}
Reader identity was never exported: the mark sets are uid-free and shuffled within each document,
so a reader cannot be followed from one document to another. That is the property the artifact was
built for, and it is also precisely what would be needed to test the assumption underneath our
inference. We therefore report \textbf{\corpMarkSets{} reader mark sets}, not \corpMarkSets{} people --- a
core of heavy users could account for many of them and nothing on disk would distinguish that from
\corpMarkSets{} individuals.

The consequence is specific. Our bootstrap clusters on domain; if the same readers recur across
documents, per-document values are correlated through readers as well, and the human side's
intervals are narrower than they should be. The direction is known and the magnitude is not. It
does not touch the ordering, which the sign test establishes on document counts alone --- model
agreement exceeds reader agreement on \signHigher{} of \signDocs{} documents, a statement no
standard-error estimate enters.

\paragraph{Readers are a selected population.} Only readers with at least $b$ marks enter, which is
\corpSelectionPct\% of reader--document pairs, selecting toward heavier markers. One task, one corpus, one
platform.

\paragraph{Some small models did not do the task on every document.} \degenArmsAny{} of the
\nArms{} arms return the sentences in document order on at least one document --- an identity
ranking, which is naive truncation. \degenArmsOften{} do so often enough to matter: Llama 3.1 8B on
\degenLlamaEight{} of \degenLlamaEightN{} documents (\degenLlamaEightPct\%), and Nemotron, Phi-4 and
Granite on between \degenOftenMinPct\% and \degenOftenMaxPct\%. The rest are occasional, and
\degenFrontierN{} of them are frontier arms, at up to \degenFrontierMaxPct\% of their documents ---
so this is not purely a small-model failure, though it is overwhelmingly one. No arm's ranking
correlates with document order above \degenMaxCorr. The estimator absorbs these: an identity
ranking's top-$k$ is the lead block, which the depth-matched null neutralises, and the algorithm
control confirms it --- naive truncation agrees with everything at approximately zero. They are
reported because an arm that silently failed a fraction of its documents is not the same object as
an arm that answered them, and a reader comparing small models to frontier ones should know which
they are looking at.

\paragraph{Excluded arms.} Three arms fell below the pre-registered coverage floor and are excluded
from pooled statistics and named: Gemini 2.5 Pro (no longer served), OLMo-3 (no endpoint), Jamba
Large 1.7 (\jambaCov/\corpDocs{} parseable, and document order on \degenJamba{} of those, which is
a second reason to exclude it). Phi-4 initially failed all \corpDocs{} documents through a defect in
our harness --- we requested more output tokens than its context window --- which is reported
because a harness defect misreported as a model property is the same failure class this work exists
to avoid.

\section{Data and ethics}

The corpus is public web documents together with mark index sets from readers of a social
highlighting platform, one of whose authors operates it. No user identifier, no highlight text, no
URL and no per-reader timestamp appears in any artifact we release: mark sets are integer sentence
indices, shuffled within each document so a reader cannot be followed across documents, and mark
times are released only as per-document month histograms. The measurement is aggregate throughout;
no individual reader is described, ranked or characterised anywhere in this paper, and the analysis
required no contact with users.

Readers agreed to public sharing of their highlights as the platform's core function, and the
documents are public pages. We nonetheless treat the reader-level data as sensitive because the
combination of a person and what they chose to mark is revealing even when each half is public,
which is why the released artifacts carry indices rather than text.

\section{Conclusion}

Measured against readers who were never given a task, language models converge on each other far
faster than they converge on people. The convergence is strong enough that a rival laboratory's
model can sit closer to a frontier model than that laboratory's own previous model sits to itself,
and it is graded: the smallest models in our panel are no more alike than two readers are.

The trend is the part we did not expect and would have missed by asserting it, and the part our
design constrains most: generation is confounded with weight regime and scale in this panel, so we
report it as suggestive. Newer models do read more like a person --- agreement with readers rises
across generations at \trMrFactor$\times$, against \trMmFactor$\times$ for agreement between models. But the two have different amounts of room left, and we had to
measure the ceiling to see it rather than argue for one --- our first argument for a bound was
wrong, because a predictor aimed at a crowd's centroid beats any two of its members. Against the
consensus of the other readers, the best of \nArms{} arms has taken \armMaxShare\% of what is
achievable on the reader side, and the median arm \armMedianShare\%. Between models the same
ceiling is exceeded \ceilModelOver{} times over.

One number in this paper is not what it looks like, and we say so where it appears: the size of
the gap depends on a procedural choice. A model's set is the sharpest $b$ it can name and a
reader's is a random $b$ of what they marked, and putting the models through the readers'
procedure halves the gap. It does not close it at any level of over-marking where either side is
distinguishable from chance; at the heaviest dilution neither side is, and the comparison stops
being informative rather than reversing. The ordering
is the finding; the multiple is a measurement of the comparison we ran.

What separates the two sets is not register: at equal depth and equal length, no surface feature we
measured distinguishes what models choose from what readers choose. Models and readers select
sentences that look the same and are not the same, and the distance between them is growing on one
side only.

\end{document}